\documentclass[reprint,apl,aip]{revtex4-1}

\usepackage{graphicx}
\usepackage{bm}

\renewcommand{\vec}{\bm}

\begin{document}

\title{
Nonuniform Spin-Wave Softening in Two-Dimensional Magnonic Crystals as a Tool for Opening Omnidirectional Magnonic Band Gaps
}

\author{S.~Mamica}
	\email{mamica@amu.edu.pl}
	\affiliation{Faculty of Physics, Adam Mickiewicz University in Pozna\'n, ul.~Umultowska~85, 61-614 Pozna\'n, Poland}

\author{M.~Krawczyk}
	\affiliation{Faculty of Physics, Adam Mickiewicz University in Pozna\'n, ul.~Umultowska~85, 61-614 Pozna\'n, Poland}

\author{D.~Grundler}
	\affiliation{Ecole Polytechnique F\'ed\'erale de Lausanne, School of Engineering, Institute of Materials and Institute of Microengineering, Laboratory of Nanoscale Magnetic Materials and Magnonics, 1015 Lausanne, Switzerland}

\date{\today}

\begin{abstract}
By means of the plane wave method we study spin wave dynamics in two-dimensional bicomponent magnonic crystals based on a squeezed hexagonal lattice and consist of a permalloy thin film with cobalt inclusions. We explore the dependence of a spin wave frequency on the external magnetic field, especially in weak fields where the mode softening takes place. For considered structures, the mode softening proves to be highly nonuniform on both the mode number and the wave vector. We found this effect to be responsible for the omnidirectional band gap opening. Moreover, we show that the enhancement of the demagnetizing field caused by the squeezing of the structure is of crucial importance for the nonuniform mode softening. This allows us to employ this mechanism to design magnonic gaps with different sensitivity for the tiny change of the external field. The effects we have found should be useful in designing and optimization of spin wave filters highly tunable by an external magnetic field.\end{abstract}

% ...

\maketitle

\section{Introduction}\label{sec_intro}

The variation of an external magnetic field applied to a magnetic system leads to a shift of the spin-wave spectrum on the frequency scale. It has already been found that at low fields the shift is nonuniform \cite{Topp_gap, Tacchi_gap, Montoncello_soft, Langer}, especially a softening of modes, i.e. decreasing of their frequencies close to zero. The nonuniformity reflects two different effects: different frequency shifts for different modes and/or a wave-vector-dependent shift within the single mode. In this paper, we explore both effects in two-dimensional (2D) magnonic crystals (MCs) \cite{Vasseur_PWM, Nikitov_MCs} attributing them to the spin-wave amplitude distribution combined with the growing influence of the demagnetizing field at low magnetic fields. In particular, we study the role of these effects in the occurrence of forbidden frequency ranges (band gaps or stop bands) in the spin-wave spectrum.

The existence of magnonic band gaps has already been reported in the literature although most of them are partial (directional) gaps \cite{Kostylev, Ma_gap, Tacchi_gap_2D, Tacchi_2DMC, Mamica_mFT, Klos_gap,  Kumar_gap, Rychly_gap, Di_gap}. As in other periodic composites, the dispersion relation of MCs can be tailored by adjusting the structure and material composition which allows manipulating the velocity, the direction of propagating spin waves, and the magnonic gap width \cite{Krawczyk_SW, Vasiliev_SW, Wang_SW, Serga_SW, Lenk_SW, Tkachenko_SW, Mandal_SW, Rychly_SW}. However, the gap tailoring during operation required a high magnetic field (of up to 1.0~T) or magnetization reversal \cite{Ma_gap, Schwarze}. 

The aim of our current work is to unfold new mechanism--based on the nonuniform mode softening--responsible for the opening of magnonic band gaps and to employ this mechanism and design band structures of the 2D MCs with gaps reversibly tunable by a tiny change of the external magnetic field magnitude (50--200~mT). We propose 2D MCs with such properties, which are feasible to be realized by the state-of-the-art technology and for which the spin-wave spectrum exhibits field-dependent omnidirectional (complete) band gaps with different sensitivity of the gap width to the external field magnitude. The proposed mechanism leads to the opening of gaps in the low-frequency range for in-plain magnetized bicomponent 2D MCs even for small magnetic contrasts of constituent materials. This makes an important step-forward in comparison with previous studies of the complete gap existence in different 2D systems, such as arrays of ferromagnetic dots, antidots lattices or bicomponent structures.

\section{Model}

\begin{figure}
\includegraphics{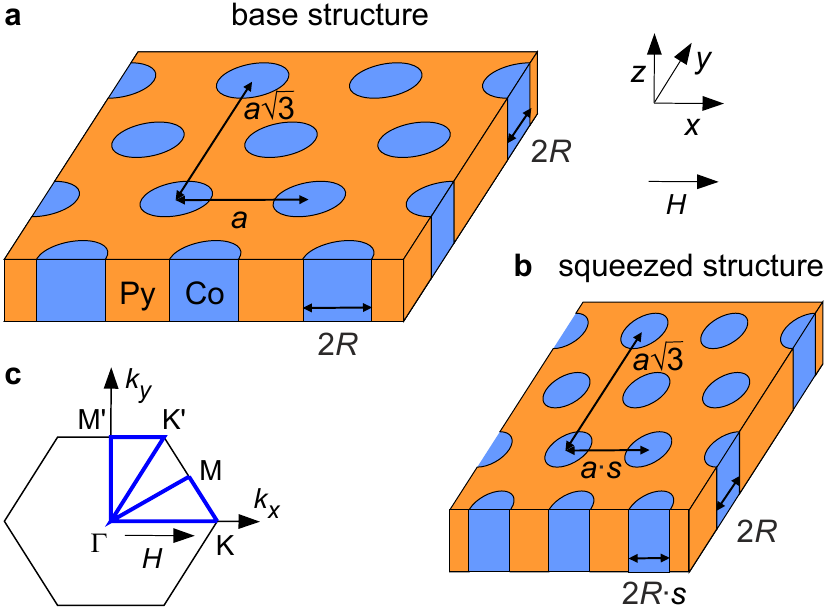}
\caption{Thin-film MC based on the 2D hexagonal lattice composed of cobalt rods (blue) in a permalloy matrix (orange). (a) The base structure with the lattice constant $a$ and the radius of Co rods $R$. (b) The structure squeezed in the $x$ direction by the structure ratio $s$. (c) First Brillouin zone for the base structure shown in (a) with high-symmetry paths indicated (blue lines).}
\label{Fig1}
\end{figure}

The system under consideration is schematically shown in Fig.~\ref{Fig1}. In panel (a) we show the base structure, which is in the shape of thin permalloy (Py) film with an array of cylindrical cobalt inclusions (rods). For the following study, we consider parameters of the structure consistent with state-of-the-art bicomponent MCs \cite{Tacchi_2DMC, Duerr_APLMC}. Rods consisting of Co are arranged in the sites of a hexagonal lattice with the lattice constant $a = 600$ nm. The diameter of the rods is 340 nm and the film thickness is 30 nm. An external magnetic field $H$ is applied in the plane of the MC along the $x$ direction. In the study, we consider MCs where the base structure is `squeezed' in the direction of the external field (Fig.~\ref{Fig1}b). We describe the squeezed structure by the ratio of the new lattice constant in $x$ direction compared to the original one which we refer to as the structure ratio ($s$). In Fig.~\ref{Fig1}c, we provide the first Brillouin zone (FBZ) for the base structure; squeezing of the structure leads to the elongation of the FBZ in the $k_{x}$ direction. Blue lines mark high-symmetry paths in the FBZ along which dispersion relations are evaluated.

We use the plane-wave method \cite{Vasseur_PWM} (PWM) to calculate spin-wave frequencies and their profiles as well as the demagnetizing field. The method is based on the linearized damping-free Landau-Lifshitz equation with assumed full saturation of the magnetization. As the considered structure is assumed to be periodic in the plane of the Py film, the material parameters can be Fourier expanded. Bloch's theorem applies to the dynamic functions, such as the dynamic demagnetizing field components and the dynamic components of the magnetization. The final set of algebraic equations is solved by the numerical diagonalization. The approach suitable for thin film bicomponent MCs is described in Refs.~\onlinecite{Sokolovskyy_PWM, Krawczyk_PWM}. Material parameters used in this work are as follows: the saturation magnetization, $M_{S}$, 1.39e6 A/m for Co, and 0.81e6 A/m for Ni80Fe20 (Py), the exchange stiffness constant 2.8e-11 J/m in Co, and 1.1e-11 J/m in Py. In the expansion, we use 271 plane waves, a number large enough to ensure the satisfactory convergence of the results.

We introduce a concentration factor, which for rods reads \cite{Mamica_mFT}:
\begin{equation}\label{eq_cf}
 cf_{A} = \frac{ \tilde{m}_{A} }{ \tilde{m}_{A} + \tilde{m}_{B} } .
\end{equation}
In the 2D case $\tilde{m}_{X} = \frac{1}{S_{X}} \int_{S_{X}} |\vec{m}|^{2} dS $ is the mean value of the squared amplitude of the dynamic magnetization in the area $S_{X}$, that is, in the rods (for $X = A$) or in the matrix ($X = B$). The quantity given by Eq.~(\ref{eq_cf}) allows us to determine in which material any particular spin-wave mode is excited mostly. By this definition a concentration factor value above 0.5 means that the concentration of dynamic magnetization is higher in Co than in Py.

\section{Results and Discussion}

\begin{figure*}
\includegraphics{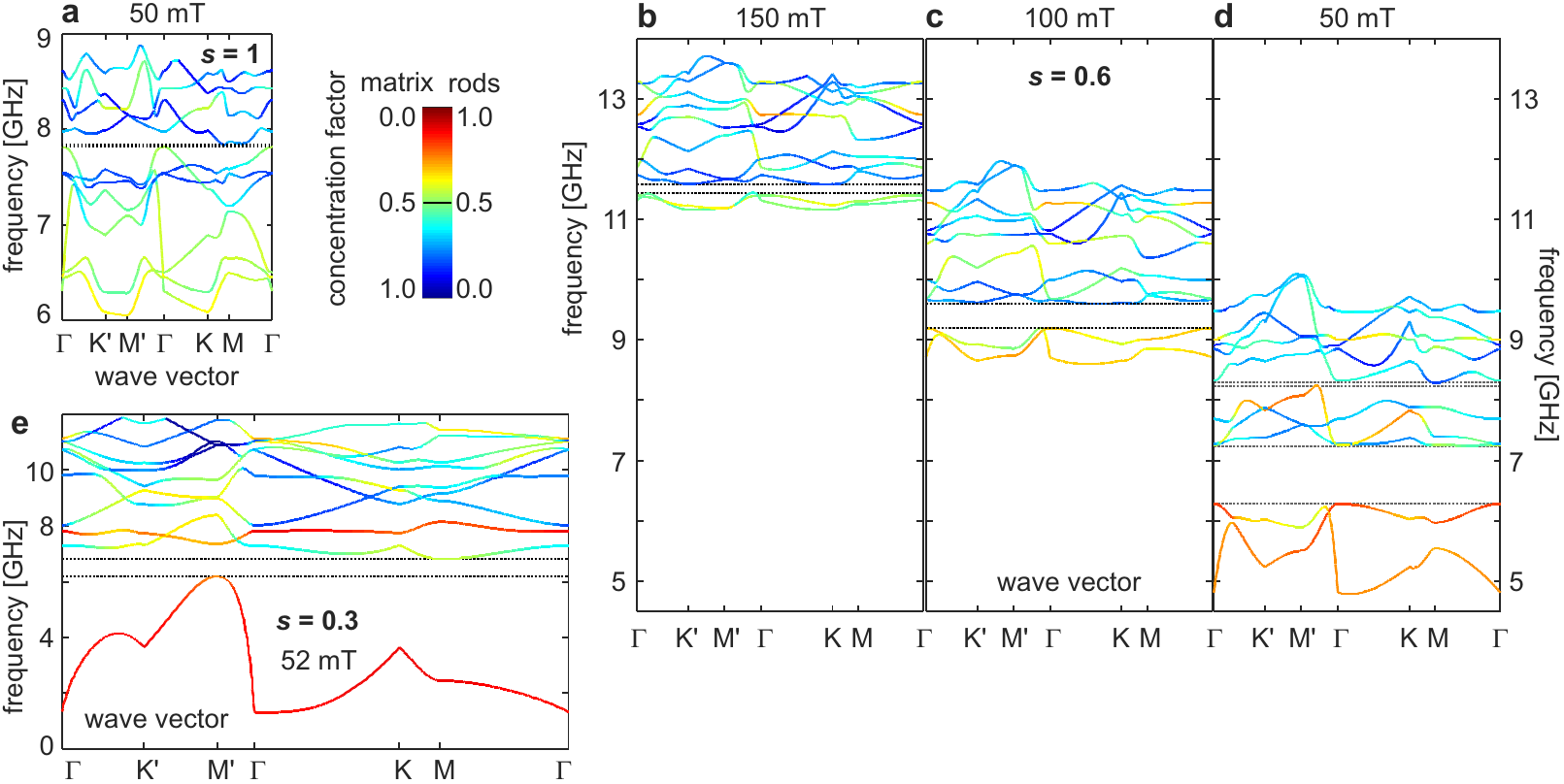}
\caption{ Spin-wave spectra of Co/Py 2D MCs along the high-symmetry paths in the FBZ (compare Fig.~\ref{Fig1}c) for (a) the base structure at 50 mT, and the squeezed structure with a structure ratio $s = 0.6$ at (b) 150, (c) 100, and (d) 50 mT. (e) The spectrum for the squeezed structure with $s = 0.3$ at 52 mT. Line colors depict the concentration factor calculated from Eq.~(\ref{eq_cf}) according to the color scale shown in the inset of (a). Dotted horizontal lines represent the upper and lower band edges of complete magnonic gaps.}
\label{Fig2}
\end{figure*}

In Fig.~\ref{Fig2} we show spin-wave band structures calculated along paths in the FBZ shown in Fig.~\ref{Fig1}c. In all graphs, only the ten lowest bands are shown and colored according to their concentration-factor value in rods or matrix (color scale is given in the inset in Fig.~\ref{Fig2}a). Dotted horizontal lines stand for frequencies limiting complete magnonic gaps. Figure \ref{Fig2}a  shows the spectrum for the base structure at the external field of 50 mT. The ten lowest branches fit the frequency range from 6 to 9 GHz. There is a very narrow (about 22 MHz) omnidirectional band gap just below 8 GHz. Figures \ref{Fig2}b, c and d stand for the structure ratio $s = 0.6$ at 150, 100, and 50 mT, respectively. In panels a, d, e, and f, the field is similar while the structure ratio changes.

In the following, we discuss the changes in the spin-wave band structures while the external field is reduced. Comparing Figs.~\ref{Fig2}b and d for $s = 0.6$ a smaller external field brings smaller spin-wave frequencies, but additionally, it leads to a broadening of the complete magnonic band gap and widening of the bandwidths of some minibands. The direct cause for the band gap broadening is a different sensitivity of the softening for different modes: frequencies of two lowest modes (minibands) go down much faster than others. Among the high-frequency modes (above the band gap) there is one that is strongly concentrated in rods (especially at 50 mT). The frequency of this mode falls down faster with decreasing field than the others. This behavior results in the opening of another gap around 8.3 GHz for 50 mT. We address this feature to the growing importance of the demagnetizing field while the external field gets weaker.

\begin{figure}
\includegraphics{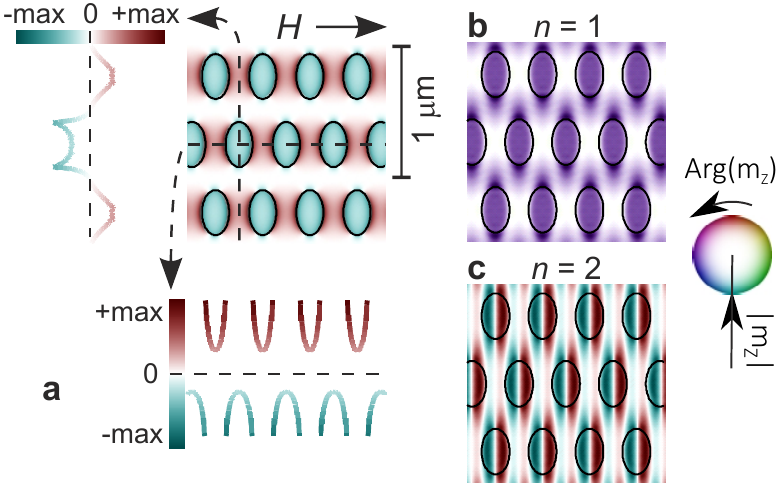}
\caption{(a) Demagnetizing field $H_{d}$ for the squeezed Co/Py MC ($s = 0.6$) along with its cross-sectional profiles parallel (bottom) and perpendicular (left) to the external field. (b, c) Spin-wave profiles for the MC in question in an external field of 50 mT for the lowest ($n = 1$) and the second ($n = 2$) mode in the spin-wave spectrum at the center of the FBZ (compare Fig.~\ref{Fig2}d). Colors represent argument (phase) and their intensity the modulus of the dynamic magnetization, as shown in the inset.}
\label{Fig3}
\end{figure}

The relative variation of the demagnetizing field for the structure ratio $s = 0.6$ is given in Fig.~\ref{Fig3}a. In the Py matrix, the demagnetizing field $H_{d}$ is large and positive between neighboring rods along the direction of the external field (bottom panel in Fig.~\ref{Fig3}a). In the Co rods, the demagnetizing field decreases internal field with deep minima near their borders. (Here `internal field' means the sum of the external magnetic field and the demagnetizing field $H_{int} = H + H_{d}$ \cite{book_Demokritov}.) Thus the modes that are concentrated in the rods effectively experience a lower internal magnetic field $H_{int}$ than those concentrated in the matrix. We can infer, that the mode softening depends on where the spin-wave amplitude is concentrated, i.e., inside or between the rods.

Indeed, we find that the resonant spin-precessional motion of modes concentrated in Co occurs at a lower frequency compared to modes in Py (Fig.~\ref{Fig2}). In unpatterned thin films of Co and Py, this would not be the case. The resonance frequency in Co is much larger than in Py due to its 70\% larger saturation magnetization $M_{S}$. $M_{S}$ and the internal field $H_{int}$ enter the equation of motion, and resonant spin precession shifts to higher frequencies with both $M_{S}$ and $H_{int}$. Due to $H_{d} < 0$ frequencies of modes concentrated in Co fall below Py (Fig.~\ref{Fig3}a). In the following, we show how this `band inversion' concerning spin precession in Co and Py can be optimized to achieve complete band gaps in squeezed hexagonal MCs.

The strong negative demagnetizing field results in a pronounced `softening' of modes concentrated in Co rods at low external fields and thus should lead to three effects. First, as described above, the lowering of frequency due to softening depends on the concentration factor. Second, the relevant modes show an increased concentration inside rods at small fields (reflected by the different colors of bands at different external fields in Figs.~\ref{Fig2}b-d). Increasing concentration means that the spin wave prefers to be excited inside rods rather than in the matrix while the external field decreases. Third, extreme softening of modes should induce magnetization reversal, which should start from the cobalt rods if the magnetocrystalline anisotropy is absent. The analysis of the reversal is out of the scope of the paper, but we note that the spin-precessional motion of the softest mode is strongly concentrated in Co rods (see Fig.~\ref{Fig3}b), which suggests the starting regions of the magnetic reorientation \cite{Mamica_vortex}, according to the nucleation field theory \cite{book_Aharoni}.

Besides the fast softening with decreasing external field one can also observe a widening of the bandwidths of minibands related to modes strongly concentrated in rods (Fig.~\ref{Fig2}b-d). This means that the softening of such modes depends on the wave vector $\vec{k}$; it appears to be anisotropic and nonuniform in $\vec{k}$ space. The possible origin of this effect is discussed later. The anisotropic softening was described for the lattice of interacting magnetic dots in Ref.~\onlinecite{Montoncello_soft}. In that case, a propagating mode was found to be a nucleation mode while in our work the lowest frequency stands for the FBZ center.

In Fig.~\ref{Fig2}, we see, that the important factor in proposed magnonic band-gap-opening mechanism is also a squeezing of the structure along the magnetic field direction. Stronger squeezing of the lattice brings two coexisting effects. First, there is less space for both types of excitations concentrated in rods and in the matrix along the squeezing direction. As in the case of the confinement of electrons in a potential well, we can expect frequencies of spin waves to shift up due to this effect. Second, the absolute value of the demagnetizing field of the negative sign inside rods and of the positive sign in the matrix along the line between the nearest rods is found to increase with decreasing $s$. In the middle of the rods the demagnetizing field amounts to -49 mT for $s = 0.6$ and -88 mT for $s = 0.3$. This feature results in a stronger concentration of spin-precessional motion in the rods (for $n = 1$ in the FBZ center the concentration factor is 0.734 for $s = 0.6$ and 0.853 for $s = 0.3$). 

In Figs.~\ref{Fig3}b and c the profiles of the two lowest modes are presented, both taken in the FBZ center for $s = 0.6$. The lowest one ($n = 1$) is the so-called fundamental mode, a counterpart of universal excitation \cite{Mamica_fundametal}, with the magnetization precession all in phase. The second one ($n = 2$) exhibits one nodal line in the middle of the rod along its longer axis. This change of phase within the rod makes the mode frequency very sensitive to the spatial confinement and thus `protects' it from softening caused by the growing of the demagnetizing field. Indeed, its frequency slightly increases with squeezing (Figs.~\ref{Fig2}d, e), which leads to a closing of the band gap existing near 7 GHz at 50 mT in Fig.~\ref{Fig2}d, i.e.,  between the 2nd and 3rd band. The phase change within the rod depends on the wave vector, so does the `protection' from softening. This feature explains the separation of the two lowest modes and, in consequence, the opening of the additional frequency gap near 6.5 GHz for $s = 0.3$ in Fig.~\ref{Fig2}e between 1st and 2nd band. The fast softening with decreasing $s$ causes the appearance of a zero-frequency mode for $s = 0.3$ at 50 mT, which might induce reversal as discussed above. To avoid ambiguities and ensure a finite frequency for the lowest miniband, we show the band structure for $s = 0.3$ at a slightly larger field of 52 mT in Fig.~\ref{Fig2}e.

\begin{figure*}
\includegraphics{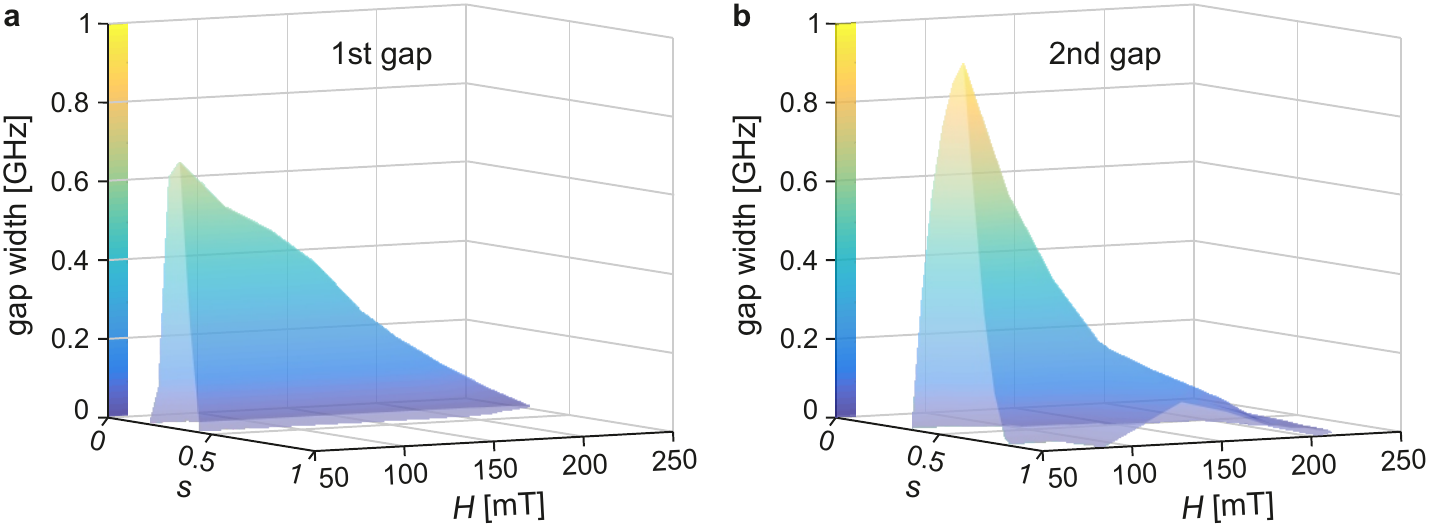}
\caption{Width of magnonic band gaps vs. structure ratio $s$ and external field magnitude $H$ for (a) the first and (b) the second gap. Colors depict the gap width.}
\label{Fig4_3D}
\end{figure*}

\begin{figure}
\includegraphics{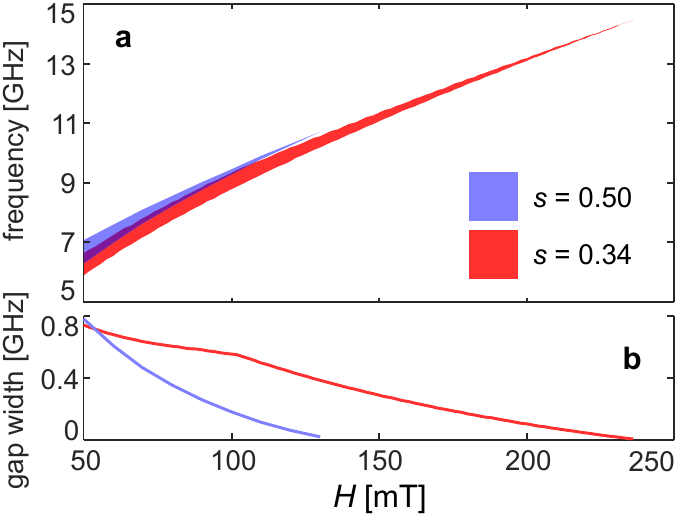}
\caption{(a) Magnonic band gaps vs. the external field magnitude $H$ for two structure ratios: $s = 0.5$ and $s = 0.34$. (b) The dependence of the gap width on $H$ for both gaps shown in (a).}
\label{Fig4}
\end{figure}

In Fig.~\ref{Fig4_3D} we show the width of two lowest magnonic band gaps observed in magnonic spectra in Fig.~\ref{Fig2} as a function of both factors: the external magnetic field magnitude and the structure ratio. For both gaps the color scale is the same and encodes their width. The first gap (Fig.~\ref{Fig4_3D}a) occurs for smaller $s$ than the second one, i.e. for more squeezed structures. For each field the  maximum appears near $s \approx 0.35$. At 50 mT the first gap amounts to about 0.8 GHz. This gap closes at 240 mT. The second gap (Fig.~\ref{Fig4_3D}b) appears for $s > 0.35$ and the position of its maximum width shifts from $s = 0.6$ to 0.65 while the external field grows from 50 mT to 150 mT. For a given $s$ the width depends on $H$ more strongly than the width of the first gap. To compare field dependence of these two gaps in detail we show the magnonic band-gap evolution with the change of the external magnetic field magnitude for two exemplary structure ratios in Fig.~\ref{Fig4}a: $s = 0.34$ for the first gap and $s = 0.5$ for the second gap. The colored regions indicate the absolute frequencies covered by the forbidden frequency gaps. In Fig.~\ref{Fig4}b the gap width dependence on $H$ is given. The gaps in both structures occur for a similar frequency range and exhibit almost the same width at 50 mT ($\approx$ 0.8 GHz). Both gaps  move to higher frequencies while $H$ increases. However, the width of the gap for $s = 0.5$ is reduced much faster with increasing $H$ than for $s = 0.34$, and the gap closes already at around 130 mT. For $s = 0.34$ the gap exists up to almost 240 mT. Thus we can propose structures based on the squeezed hexagonal MCs with a different sensitivity of the gap width to the external field magnitude.

\section{Conclusions}

In conclusion, we show that at the low magnetic field the demagnetizing field proves to be responsible for complete band gaps in the spin-wave spectrum in squeezed hexagonal MCs. Its growing importance with decreasing external field causes the mode softening to be strongly dependent on the concentration of dynamic magnetization in rods as well as on the wave vector. The $\vec{k}$-dependent softening leads to the broadening of minibands. Tailoring of gaps is possible via demagnetizing field design. Their absolute values are controlled by the external magnetic field. Different softening for different modes causes complete magnonic gaps to open and close just by changes of the external magnetic field magnitude. Proposed structures, originating from the hexagonal lattice of Co rods immersed in the Py matrix, could be fabricated with the current technology, and the band gap can be determined in Brillouin-light-scattering experiments or in the transmission measurements with VNA-FMR \cite{Krawczyk_Grundler}. The reversible control of omnidirectional band gaps in 2D MCs can make them very useful for designing of the tunable spin wave filters and transducers.


\begin{acknowledgments}
The study has received financial support from the EU's Horizon 2020 research and innovation programme under Marie Sklodowska-Curie GA No. 644348 (MagIC), from the Polish Ministry of Science and Higher Education resources for science in 2017--2019 granted for the realization of an international co-financed project (W28/H2020/2017), and from the National Science Centre of Poland under Grant No. UMO-2016/21/B/ST3/00452.
\end{acknowledgments}



\end{document}